# Laser-induced helicity and texture-dependent switching of nanoscale stochastic domains in a ferromagnetic film

Dinar Khusyainov[1*], Rein Liefferink[1], MengXing Na[1], Kammerbauer Fabian[2], Robert Frömter[2], Mathias Kläui[2], Dmitry Kozodaev[3], Nikolay Vovk[4], Rostislav Mikhaylovskiy[4], Dmytro Afanasiev[1], Alexey Kimel[1], Johan H. Mentink[1], Theo Rasing[1]

[1*]*Institute for Molecules and Materials, Radboud University, Nijmegen, 6525 AJ, Netherlands.*
[2]*Institute of Physics, Johannes Gutenberg University Mainz, Mainz, 55099, Germany.*
[3]*NT-MDT BV, Sutton 11A, Apeldoorn, 7327AB, Netherlands.*
[4]*Department of Physics, Lancaster University, Bailrigg, Lancaster, LA1 4YW, United Kingdom.*

*Corresponding author. E-mail: dinar.khusyainov@ru.nl;

Controlling magnetic textures at ever smaller length and time scales is of key fundamental and technological interest. Achieving nanoscale control often relies on finding an external stimulus that is able to act on that small length scales, which is highly challenging. A promising alternative is to achieve nanoscale control using the inhomogeneity of the magnetic texture itself. Using a multilayered ferromagnetic Pt/Co/Pt thin-film structure as a model system, we employ a magnetic force microscope to investigate the change in magnetic nanotextures induced by circularly polarized picosecond laser pulses. Starting from a saturated magnetic state, we find stochastic nucleation of complex nanotextured domain networks. In particular, the growth of these domains depends not only on their macroscopic magnetization but also on the complexity of the domain texture. This helicity and texture-dependent effect contrasts with the typical homogeneous growth of magnetic domains initiated by an effective magnetic field of a circularly polarized laser pulse. We corroborate our findings with a stochastic model for the nucleation of magnetic domains, in which the nucleation and annihilation probability not only depends on the helicity of light but also on the relative magnetization orientation of neighboring domains. Our results establish a new approach to investigate ultrafast nanoscale magnetism and photo-excitation across first-order phase transitions.

## Introduction

Femtosecond laser pulses have led to a plethora of new and unexpected realizations of phase transitions between different crystallographic [1–6], electronic [7–11], or magnetic order [12–19] that can be controlled on ultrafast timescales. While these phase transitions have been extensively studied, it remains challenging to discern how ultrafast phase transitions occur at the nanoscale. In this regard, ultrafast phase transitions characterized as "first-order" in equilibrium are particularly interesting. Examples include the metal-to-insulator transition [3, 20, 21], reversal of ferroelectric polarization [22], as well as transition to a hidden electronic state [23, 24]. These first-order phase transitions feature hysteresis and phase coexistence. Therefore, nanoscale (meta)stable domains of one phase can be photo-induced inside the pre-existing stable phase on an ultrafast time scale. Recent studies indicate that the inherent stochasticity of the domain nucleation is essential for understanding the transition pathway [3, 25]. Additionally, recent studies on laser-induced topological phase transitions involving the stochastic formation of nanoscale skyrmions [14] were shown to rely on fluctuations. Therefore, a deeper understanding of the control of such photo-induced stochasticity is of key interest. Moreover, it may unlock conceptually new ultrafast brain-inspired computing paradigms based on memristive and probabilistic phenomena [26–32].

Distinct from structural and electronic, magnetic phase transitions offer an additional degree of freedom due to the dependence of the light-spin interaction on the helicity of light. This has been widely exploited for transparent magnetic media based on the ultrafast inverse Faraday effect [33–35]. Moreover, in metallic magnets, magnetic circular dichroism (MCD) can be utilized to control the magnetic state [36–39]. It has been suggested that the control of magnetization with circularly polarized light depends on the local magnetization or the difference in the absorption of circularly polarized light between two oppositely oriented magnetic domains. The latter can cause domain wall motion induced by a thermal gradient across the domain wall [37, 40–42]. Although these all-optical experiments were conjectured to be accompanied by the formation of sub-micrometer magnetic domains [40, 42, 43], direct experimental evidence of helicity-dependent control of nanoscale magnetic textures is missing so far [44].

Studying the effect of circularly polarized light on photo-induced magnetic nanotextures requires observing nanoscale changes after exposure to picosecond laser pulses. This is a challenging task where conventional optical microscopy fails due to the fundamental diffraction limit. Magnetic force microscopy (MFM) is a tabletop technique that can solve these challenges. MFM is a well-established and widely used technique because of its high spatial resolution (∼10 nm) and simplicity [45, 46]. Indeed, MFM has enabled the study of magnetic domain walls [47] and topological magnetic particles such as vortexes [48], skyrmions [49, 50], and even merons [51]. Moreover, MFM has also been recently employed to study nanotextures induced after all-optical switching in ferrimagnetic Tb/Fe multilayers [52].

Here, we employ MFM to study the laser-induced magnetic nanotextures in a ferromagnetic trilayer Pt/Co/Pt. Our measurements reveal that consistent with the established physics of first-order phase transitions, the first few laser pulses trigger the stochastic nucleation of sub-micrometer magnetic domains. By systematically studying the evolution of the photo-induced magnetic nanotexture of the micrometer and sub-micrometer domains as a function of the number of circularly polarized incident pulses, we find that the complexity of the nanotexture of these domains evolves differently than the total switched area. Moreover, by varying the laser fluence, we are able to distinguish between the nucleation and heating of the magnetic nanotextured domains. In particular, we demonstrate that under the action of laser radiation, isolated sub-micrometer domains shrink instead of grow. Both these findings show that existing mechanisms for all-optical magnetization switching, which rely on an effective magnetic field originating from the inverse Faraday effect [53, 54] or a temperature gradient-driven domain wall motion [40, 41, 55], are not able to describe the effect of picosecond circularly polarized laser pulses on magnetic nanotextured domains. To explain our experimental results, we developed a phenomenological model based on switching energy barriers of the nanoscale domains that account for the MCD effect and depend on the relative orientation of adjacent domains, achieving effective domain growth via nucleation and coalescence.

## Results

The sample we studied is a ferromagnetic Pt/Co/Pt trilayer with perpendicular magnetic anisotropy (as shown in Fig. 1a). We first establish the necessary conditions for laser-induced deterministic switching (DS) of magnetization in our sample using conventional magneto-optical microscopy. In our experiments, we utilize circularly polarized laser pulses with specific helicities chosen to achieve DS of the magnetization (see the Methods section). The pulse duration is set to 3 picoseconds, as this duration provides the highest efficiency for multipulse helicity-dependent control of magnetization, consistent with previous experiments conducted on Co/Pt samples [37, 40–43].

A sketch of the phase diagram of the magnetization state ($M$) after illumination with circularly polarized pulses as a function of the number of pulses ($N$) and fluence ($F$) is presented in Fig. 1b. First, we observe that above a fluence of $F_{DM} = 2.6$ mJ/cm$^2$, the sample becomes demagnetized (DM). Below this threshold, the ferromagnetic monodomain ground state (GS) shifts to the DS state through an intermediate state with partial switching as $N$ increases. Here, at least ten pulses are required to achieve DS, while adding additional pulses changes the DS state into a DM state.

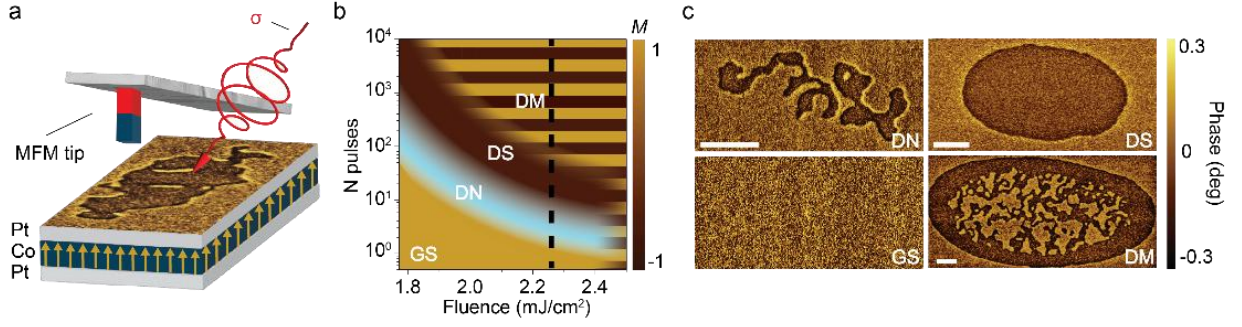

**Fig. 1 Nanoscopy of the magnetic nanotextures in Pt/Co/Pt trilayers. (a)** Sketch of the *in situ* MFM measurement of domains nucleated by picosecond circularly polarized laser pulses (σ) in Pt/Co/Pt trilayer. Arrows indicate spins that are aligned in the direction perpendicular to the surface plane. **(b)** Sketch of the phase diagram of the magnetization state *M* as a function of the number of circularly polarized laser pulses and fluence, displaying the ground state (GS), domain networks (DN), deterministic switching (DS) and demagnetized state (DM). The magnetization, denoted as $M=1$ and $M=-1$, represents the magnetization directions "up" and "down," respectively. The diagram of the laser-induced magnetization has been reconstructed using data obtained from a conventional Faraday microscopy setup (original data can be found in Extended Data Fig. 1). The dashed line indicates the fluence used to achieve DS after 10 pulses. **(c)** MFM images of the typical nanotextured domains were observed in the DM and DS state after illumination with $10^4$ pulses and DN state after illumination with 4 pulses. The scale bar is 5 μm.

To achieve nanometer-scale resolution, we aligned the laser to the scanning region of the MFM tip and performed *in situ* probing of the laser-induced nanotextured magnetic domains (Fig. 1a). Using MFM, we discovered that the intermediate partially switched state (the light-blue region in Fig. 1b) is characterized by interconnected domains, which we term domain networks (DNs). As these domain networks are typically of micrometer size, the resolution of MFM allows us to resolve the nanoscale texture of these domains. Typical MFM images displaying the magnetic domains observed in the DM, DS, and DN states are presented in Fig. 1c.

Next, we focus on the evolution of the DN under the action of different numbers of pulses. In our experiments, we use sequences of the millisecond separated pulses in the range from 0 to 10 pulses, each having a fluence $F = 2.26$ mJ/cm² (dashed black line in Fig. 1a). A representative evolution of the nucleated magnetic nanotextures as a function of *N* of circularly polarized laser pulses is shown in Fig. 2a. We repeated the experiment three times for each *N* value. After each MFM image, the irradiated region was returned to its ground state using an external magnetic field higher than the coercive field of the sample (see the Methods section). We observe that already, after two pulses, domains having a size of about 1-2 μm are generated. After four pulses, the nucleated domains coalesce into a DN state. Although the total switched area remains roughly the same in each repetition of the experiment, the locations and forms of the nucleated domains vary across the illuminated spot. These domain networks demonstrate stochastic behavior, so we refer to them as stochastic domain networks (SDN). With an increased number of pulses, we observe an increased coalescence of the SDN, resulting in nearly uniformly switched magnetization after about ten pulses.

To quantify our observations, we use two metrics. The first is the switched area ($A_s$), normalized by the total area of the laser beam ($A_b$). As shown in Fig. 2b, the growth of the $A_s/A_b$ is approximately linear in the range of 10 pulses, as indicated by the solid grey line. Secondly, we measure the complexity of the SDNs, using the fractal dimension *D* as a parameter calculated via the Minkovski-Bouligand method [56, 57], see Extended Data Fig. 2. We treat the domain wall as a one-dimensional object so that a fractal dimension $D = 1$ would refer to perfectly round domains. Fig. 2b shows the rapid growth of the domain's complexity when *N* varies from 2 to 6. One can see that during this stage, the nucleated domains begin to coalesce, forming a SDN. At 6 pulses, *D* peaks at a value of ∼1.27, where we see the most complex pattern of magnetic nanotextures in the SDN. After 6 pulses, *D* decreases to ∼1.23, indicating the formation of a DS state.

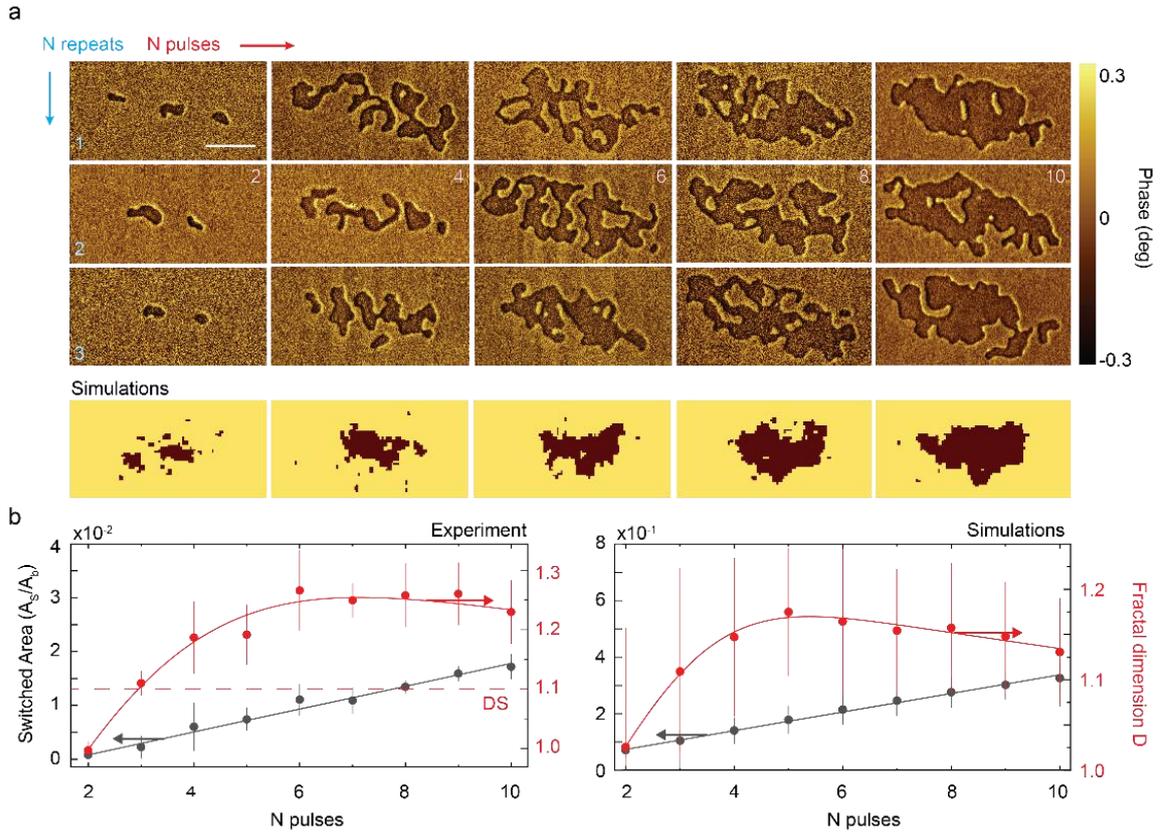

**Fig. 2 Multi-pulse evolution of stochastic domain networks**. **(a)** MFM images of the laser-induced stochastic domain networks as a function of the number of pulses $N$ at a fluence of $F = 2.26$ mJ/cm$^2$. The lower panel shows the result of the simulations using a probabilistic energy barrier-based model (details, see below). The scale bar is 5 μm. **(b)** The normalized switched area $A_s/A_b$ (gray) and the fractal dimension $D$ (red) as a function of the incident number of circularly polarized laser pulses $N$, calculated from the experiment and simulation, respectively. The solid lines serve as a guide for the eye. The switched area grows linearly up to 10 pulses, as can be seen in the experiments and simulations. In $D$, we observe a rapid increase to $D = 1.27\pm0.07$ ($1.17\pm0.08$) from 0 to 6 pulses, followed by a decrease to $D = 1.23\pm0.05$ ($1.13\pm0.06$) for experiments (simulations). A red dashed line highlights the fractal dimension in the DS state (See Fig.1 c).

The observed multi-pulse formation of the SDN deviates from the established understanding of deterministic switching in Co/Pt multilayers. The conventional models explain all-optical switching in ferromagnetic thin films as nucleation followed by the gradual growth of domains, driven by temperature differences across domain walls [37, 40–42]. This gradient arises from the MCD effect across a domain wall caused by differences in absorbed laser power for domains with opposite directions of magnetization. The subsequent temperature difference drives the domain wall from the colder toward the hotter region. For instance, temperature gradients on the order of $\nabla T = 1$ K/nm have been reported [40–42, 58]. It remains uncertain whether the temperature gradient also plays a dominant role in deterministic switching at the nanoscale. In this scenario, we would expect a steady-state growth of domains with each laser pulse, resulting in growth that is uniformly distributed along the entire domain wall. This should maintain the complexity of the nucleated domains. However, this expectation contrasts sharply with the results of our experiments.

We conducted the following experiment to investigate the contribution of the MCD-derived domain wall motion to the deterministic switching process at the nanoscale. Using the fluence $F_{DM}$, we first nucleate a set of mutually isolated domains, see Fig. 3. Next, we illuminate the nucleated domains with identical pulses but with a fluence below $F_{DM}$, thus prohibiting further nucleation of the domains. This allows us to investigate the growth of the existing domains solely. As circularly polarized pulses are more strongly absorbed by the GS state, the temperature is higher outside the sub-micrometer isolated domains, and thus, the domains are expected to grow. In contrast to these expectations, Fig. 3 shows that at $F=2.4$ mJ/cm$^2$ < $F_{DM}$, nucleated domains became smaller with each subsequent pulse until the third below-threshold pulse annihilated them. These observations prove that an effective magnetic field originating from the MCD-induced thermal gradient is not the dominant mechanism at the nanoscale. Hence, the impact of circularly polarized light on the observed sub-micrometer magnetic domains has to be attributed to another microscopic mechanism.

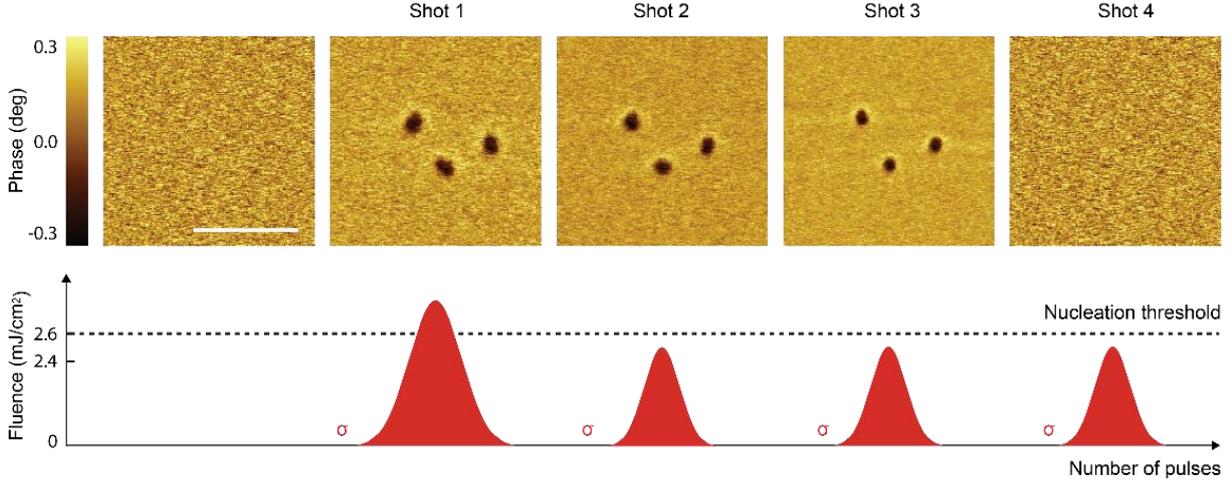

**Fig. 3 Evolution of photo-induced sub-micrometer domains under the MCD-derived temperature gradient.** First, a pulse with above threshold fluence (2.6 mJ/cm²) was used to nucleate sub-micrometer domains. The fluence of subsequent pulses was reduced to 2.4 mJ/cm² to prevent additional nucleation. We observe the shrinking and eventual annihilation of the nucleated domains. The scale bar is 5 μm.

## Discussion

To explain our experimental observations, we propose a phenomenological model that can be seen as an extension of the model for granular magnetic media developed by Gorchon *et al.* [38, 39]. To adapt this model to the case of continuous films, we consider a system composed of small domains, where each domain is defined by either magnetization up or down, denoted as $M = \pm 1$. For simplicity, we choose an equidistant mesh in two dimensions, in which each cell has the same size, corresponding to the smallest stable domain observed experimentally (see Extended Data Fig. 7). Following Gorchon *et al.*, the effect of the laser pulse is modeled as an Arrhenius activation over an energy barrier. In this model, the characteristic time for the system to transition over the energy barrier is expressed as $\tau = \tau_0 exp(E_b/k_B T)$, where $E_b$ is the energy barrier, $T$ is the electronic temperature, and $1/\tau_0$ represents the attempt frequency to overcome the barrier [38]. Distinct from granular systems, we introduce a dependence on the alignment of the adjacent domains in the switching energy barrier for continuous films. As shown in the supplementary material, it follows from micromagnetic calculations that the dependence takes the form

$$E_b(n) = E_0 + E_1 n, \qquad (1)$$

where $0 \leq n \leq 4$ is the number of aligned direct neighbors of the cell. In short, the energy barrier is reduced by the presence of domain walls with neighboring domains, leading to a linear dependence on the number of aligned neighbors. To model the dynamical changes of the domains under the influence of laser pulses, we assume that the switching probability during a short time interval $\Delta t \ll \tau$ is given by $P = \Delta t/\tau$.

We simulate the experiment by iterating over the time step $\Delta t$ (see Methods section). Ultrashort laser pulses are treated as temporal and spatial temperature profiles that act as ultrafast heating and exponential heat dissipation. The details of this are provided in the Methods section. The MCD due to circularly polarized laser pulses is implemented such that the $M = 1$ grid cells (yellow) experience a 5% higher laser heating than the $M = -1$ cells (black). The results of this simulation are shown in Fig. 2, next to the MFM measurements discussed earlier. We see that the model exhibits qualitative agreement with the experimental observations.

To better understand the physics of this model and particularly the role of MCD, we illustrate it for a simple system. Fig. 4a shows the energy barriers for a single domain as a function of the number of aligned neighbors. When no neighbors are aligned, the domain exhibits the lowest energy barrier and the highest switching probability. As the number of aligned neighbors grows, the barrier increases, and the switching probability decreases. Upon introducing the small temperature differences caused by the MCD, denoted as $\pm \Delta T$ (for $M=\pm 1$, respectively), the switching probabilities $P(n)$ are affected by $P_\sigma(n) - P(n)$, where $P_\sigma(n)$ is the probability at $T \pm \Delta T$ caused by circularly polarized light. As illustrated in Fig. 4b, the size of this change depends on the number of aligned neighbors $n$ of a given cell (at the maximum temperature reached in the simulation). To understand this point, we perform a linearization in $\Delta T/T$

$$P_\sigma(n) - P(n) \propto M \cdot exp\left(-\frac{E_0 + E_1 n}{k_B T}\right) \cdot \frac{E_0 + E_1 n}{k_B T} \cdot \frac{\Delta T}{T}. \qquad (2)$$

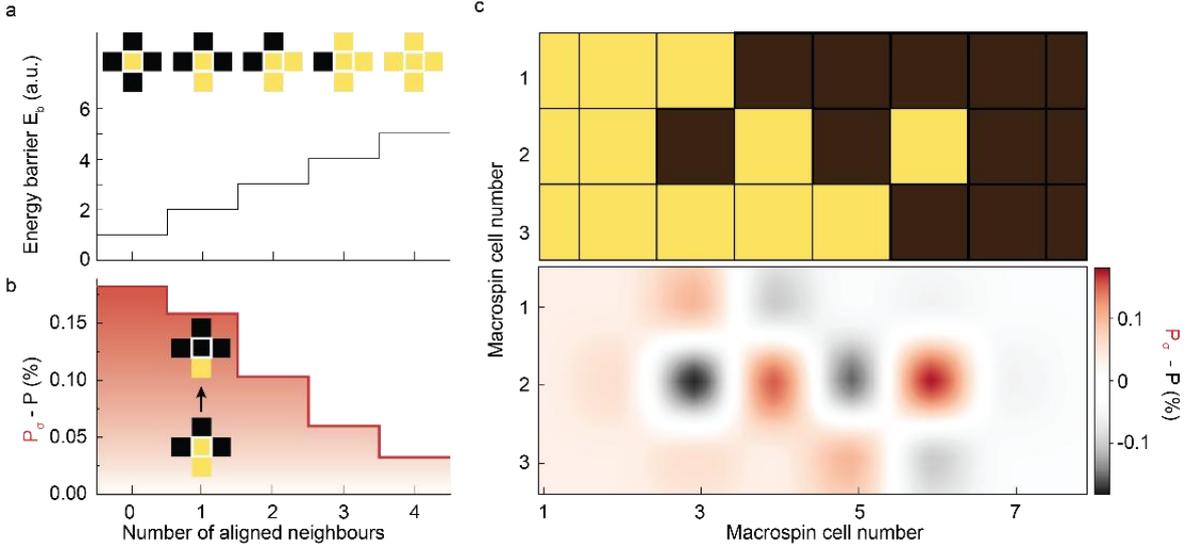

**Fig. 4 The impact of the texture and helicity of light on the switching probability**. (a) Energy barriers of the single $M=1$ (yellow) cell as a function of the aligned nearest neighbors. (b) Difference between switching probabilities with and without MCD as a function of aligned nearest neighbors at peak temperature (see Eq. (2)). The effect is maximized for zero and one aligned neighbor. (c) The macrospin model is calculated for a small grid of cells (top). The enhancement of the switching probability due to the MCD effect close to the complex domain wall is shown at the bottom.

In the simulation parameters, the condition $(E_0 + E_1 n)/(k_B T) \geq 1$ is satisfied. This ensures that the exponential decay dominates over the linear term, decreasing $P_\sigma(n) - P(n)$ as $n$ increases. This observation explains why the MCD-derived enhancement of switching probability is most significant for cells with no or one aligned neighbor. We simulate a complex magnetic nanotexture shown in Fig. 4c and calculate $P_\sigma(n) - P(n)$. The MCD's most significant effect on the switching probability is observed in the middle section of the figure, specifically for cells with $n = 0$ or $n = 1$ aligned neighbors, where the domain network's nanotexture is the most complex.

The outcome of this model is a detailed explanation of helicity and texture-dependent switching in ferromagnetic thin films at the nanoscale. Initially, a laser pulse with above-threshold fluence nucleates isolated domains stochastically. Following this, the growth of these domains occurs due to subsequent irradiation with laser pulses and nucleation of new domains and their coalescence with existing ones. The stochastic nature of this process leads to the formation of complex stochastic domain networks. When these stochastic domain networks are exposed to circularly polarized laser pulses, the switching probability is enhanced near complex magnetic nanotextures, reducing the complexity of the existing domains and forming a deterministic switched state.

## Conclusion

In summary, using *in situ* MFM measurements, we studied the impact of circularly polarized picosecond laser pulses and devised a model to explain the photo-induced nucleation and growth of magnetic nanotextured domains in a Pt/Co/Pt trilayer. Our results reveal that a single laser pulse can stochastically nucleate sub-micrometer magnetic domains. Contrary to common belief, we find that the subsequent photo-induced growth of the domains is not driven by homogeneous growth due to an effective magnetic field, as previously conjectured for sub-micrometer magnetic domains. Instead, quantitative characterization of the emerging SDN as a function of the number of laser pulses shows that the complexity of the domain network evolves distinct from the switched area. We introduced a novel measure for the switching efficiency based on the fractal dimensionality/complexity $D$, which helped us to effectively track the evolution of the SDN under irradiation of circularly polarized pulses. Our phenomenological modeling shows that the switching probability due to the MCD effect is enhanced by the complexity of the SDN nanotexture, providing qualitative agreement with the experimental results. These results suggest new pathways utilizing inhomogeneous laser-induced heating to control and investigate ultrafast nanoscale magnetism and photo-excitation across first-order phase transitions in general. These findings also suggest new opportunities to use ultrafast laser-induced SDN for probabilistic computing [26–29] and for designing memristive devices [59, 60].

# Methods

**Sample fabrication.** We engineered a multilayer structure with perpendicular magnetic anisotropy, composed of thin films of Ta (1.0 nm), MgO (2.0 nm), Pt (3.0 nm), Co (0.6 nm), Pt (3.0 nm), and Ta (4.0 nm) on a synthetic glass substrate. This structure exhibited a coercive field of approximately 28 mT, as shown in Extended Data Fig. 4. We chose to use a single Co layer based on the well-established finding that reducing the Co/Pt repeats leads to higher perpendicular magnetic anisotropy, larger domain size, and more efficient all-optical switching [43]. The symmetric structure was utilized to simplify the magnetic system and eliminate the Dzyaloshinsky-Moriya interaction [49].

**Laser excitation.** An amplified laser system with a central wavelength of 800 nm and a repetition rate of 1 kHz was used to control the sample surface's magnetization locally. We utilized 3 ps circularly polarized laser pulses with helicities defined as $\sigma^+$ and $\sigma^-$ (for right and left-handed circularly polarized laser pulses, respectively). We used a quarter-wave plate to control the helicity of circularly polarized laser pulses. To control the number of pulses during sequential illumination of the millisecond-separated pulses, we used an external Pockels cell.

**Deterministic switching conditions.** To establish the switching conditions, we utilized magneto-optical microscopy in a Faraday geometry. First, we identified the effect of the circularly polarized light pulses on the switching process using a swiping technique [36]. The result is shown in Extended Data Fig. 5. We observed that $\sigma^-$ circularly polarized pulses deterministically switch the magnetization from the "up" state ($M=1$) to the "down" state ($M=-1$). Second, we varied the fluence and the number of $\sigma^-$ pulses on different sample spots and monitored the final state of the Pt/Co/Pt exposed to laser pulses (see Extended Data Fig. 1a). We defined the condition as DS when the magnetization in the center of the spot was homogeneous, and there was no multidomain state. The switching efficiency diagram can be found in Extended Data Fig. 1b. One can see that homogeneous switching from magnetization $M = 1$ to magnetization $M = -1$ could be achieved for fluences lower than ∼2.6 mJ/cm². The minimum number of pulses required for deterministic switching was 10, consistent with previous studies [37].

**MFM.** We utilized an NT-MDT NTEGRA Atomic Force Microscope for the MFM measurements. MFM images were acquired with a probe height of less than 100 nm. Two-pass MFM method was used. Low-moment MFM tips (NT-MDT MFM LM) were utilized to avoid disturbing the magnetic domains in Pt/Co/Pt. MFM images were corrected using the median line correction method. A direct comparison of the magneto-optical snapshots with the MFM maps shows a substantial spatial resolution enhancement (see Extended Data Fig. 6). Using MFM, it was possible to observe nanoscopic domains with sizes as small as ∼200 nm (see Extended Data Fig. 7). The effect of the $\sigma^-$ light in comparison with linearly polarized light on deterministic switching was studied using MFM probing as a function of the number of optical pulses (see Extended Data Fig. 8).

**Single pulse helicity-dependent nucleation**. We quantified the effect of the single laser pulse fluence and MCD effect on the nucleation of the nanoscale domains. To this end, we performed fluence dependence measurements of both ($\sigma^-$) and ($\sigma^+$) circular polarized pulses [38–42]. The result is shown in Extended Data Fig. 3 a. We observed that the threshold for small domains is $F_c^- =2.57$ mJ/cm² for $\sigma^-$ pulses, while for $\sigma^+$ pulses, the critical threshold fluence is $F_c^+ =2.7$ mJ/cm². The difference between $F_c^+$ and $F_c^-$ confirms that the absorption of $\sigma^+$ and $\sigma^-$ pulses is inequivalent, which can be attributed to the MCD effect. In the present case, the wavelength of the laser is 800 nm, for which the typical value of the MCD in Pt/Co/Pt is approximately 1.5% [61]. In Extended Data Fig. 3 b, we plot $A_l/A_b$ as a function of helicity and fluence and extract an MCD-induced shift in the fluence dependence of approximately 3%. Extended Data Fig. 3 c shows that the fractal dimension $D$ also exhibits an MCD effect but with a slightly smaller shift of approximately 1.5%. We observe that both $A_l/A_b$ and $D$ exhibit nonlinear growth as a function of fluence. Interestingly, $D$ increases and saturates at approximately 1.4 for a 2.9 mJ/cm2 fluence.

**Simulations.** Simulations were conducted using a square grid of 100x50 cells model system. Each cell interacts with its neighbors, and the energy required to switch these domains depends on the nearest-neighbor alignments. We utilized Equation (1) to estimate the energy barriers using $E_0$ and $E_1$, which are extracted from the micromagnetic model (details are provided in Supplementary Material). The laser pulse is modeled as a temperature profile in time. For simplicity, the temperature $T(t)$ follows an exponential rise and decay defined by $dT/dt$, during the laser pulse

$$\frac{dT}{dt} = \begin{cases} (T - T_\text{p})/\tau_{rise}, & \text{during pulse} \\ (T - T_0)/\tau_{fall}, & \text{otherwise} \end{cases}. \qquad (3)$$

For the duration $w_{\text{pulse}} = 30\tau_0$ of the laser pulses, the temperature goes to a peak $T_\text{p}$ on a timescale $\tau_{rise} = \tau_0$ (for simplicity, the heating is almost instant) and to a base temperature $T_0 = 0.2 E_0/k_\text{B}$ on a timescale $\tau_{fall} = 30\tau_0$ otherwise.

The peak temperature during the laser pulse is influenced by both the position and the direction of the local magnetization. This is because the laser has a Gaussian profile in space, and the heating of the cells is affected by their magnetization $M$ through the MCD effect. This leads to

$$T_p = T_p(M, \text{r}) = T_0 + (1 - \alpha M) T_{peak} \exp\left(\frac{\text{r} - r_0}{2\sigma_p}\right), \qquad (4)$$

where $\alpha = 0.025$ is the MCD factor, leading to a 5% difference in heating, $r_0$ is the position of the laser, $\sigma_\text{p} = 8\mu m$ (40 cells) is the width of the Gaussian profile, $T_{\text{peak}} = E_0/k_\text{B}$. The time interval between laser pulses was $300\tau_0$.

Calculating the fractal dimension directly from the simulation images poses a challenge due to the low resolution, where the relevant shapes are approximately of the same size as the pixels. To effectively compare the simulations with the MFM images, we refined the mesh by a factor of 10 and applied a Gaussian filter with a standard deviation of $\sigma = 100$nm to simulate the resolution of MFM. In the box-counting method for calculating the fractal dimension, we only consider the points corresponding to the largest five box sizes for the fit (See Extended Data Fig. 2). The simulated switched area and fractal dimension data shown in Fig. 2b are the average and standard deviation over 50 runs for improved accuracy and to account for the stochastic nature of the model.

## Acknowledgments

The authors thank N. Kiselev, D. Grundler, R. Medapalli, and L. Körber for fruitful discussions, and K. Saeedi and C. Berkhout for their technical support. Additionally, the authors would like to thank N. Khokhlov, V. Bilyk, L. Nowac, and J. Hintermayr, for their assistance with the building of the experimental setup. This work was funded by the Netherlands Organization for Scientific Research (NWO), VIDI project no. 223.157 (CHASEMAG), ERC grant 485 101078206 ASTRAL, the European Union Horizon 2020 and innovation program under the European Research Council ERC Grant Agreement No. 856538 (3D-MAGiC), the European Union Horizon 2020 innovation program under the Marie Skłodowska-Curie Grant Agreement No. 861300 (COMRAD), the Gravitation program of the Dutch Ministry of Education, Culture and Science (OCW) under the research program "Materials for the Quantum Age" (QuMat) registration number 024.005.006. F.Kammerbauer and M. Kläui acknowledge funding by the Deutsche Forschungsgemeinschaft (DFG, German Research Foundation) TRR 173/2 Spin+X (Project A01 and B02 ). M. Na acknowledges the support of the Natural Sciences and Engineering Research Council of Canada (NSERC) PDF fellowship.

## Author contributions

D.A., J. H. M., A.K., and T.R. conceived the project. D.A., J. H. M., A.K., and T.R. supervised the study. D.K. performed the experiments. F.K., R.F., and M.K. fabricated the samples. D.K. analyzed and interpreted the results of MFM and magneto-optical experiments. D.K., M.N. prepared figures. R.L. and J. H. M. provided a theoretical model. D.K., M. N., J. H. M., D.A., A.K., and T.R. wrote the original draft with feedback from all the co-authors.

## Competing interests

The authors declare no competing interests.

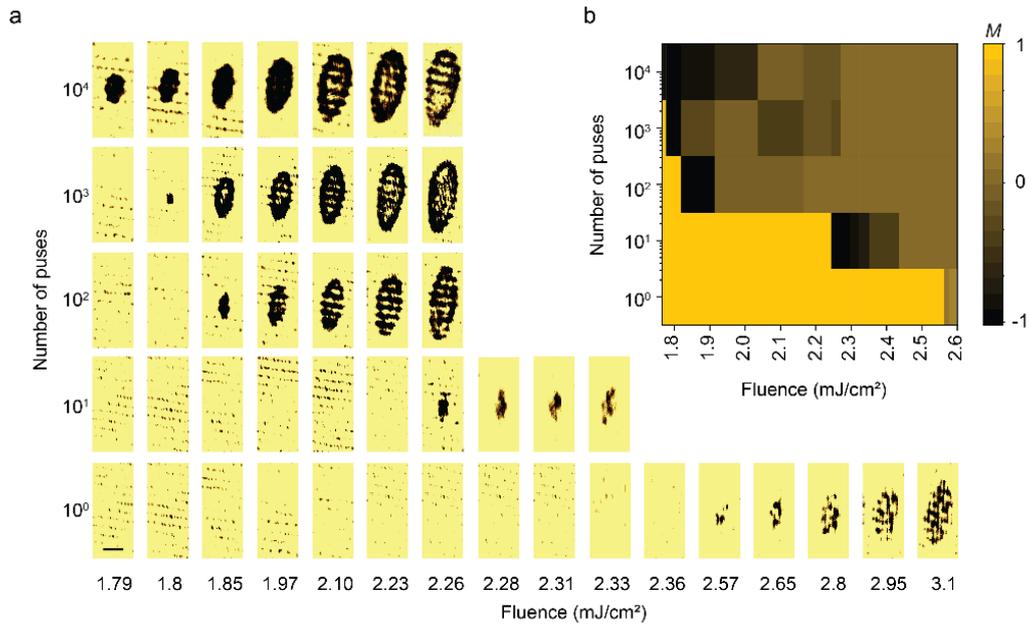

**Extended Data Fig. 1 Conditions for deterministic switching in Pt/Co/Pt.** **(a)** Magneto-optical images of the magnetization after illumination with σ laser pulses as a function of *N* and *F*. Magneto-optical images were obtained using magneto-optical Faraday microscopy. Images were post-processed by removing the background. The scale bar is 15 μm. **(b)** Diagram of the average magnetization after laser illumination inside of the switched spot as a function of *N* and *F*.

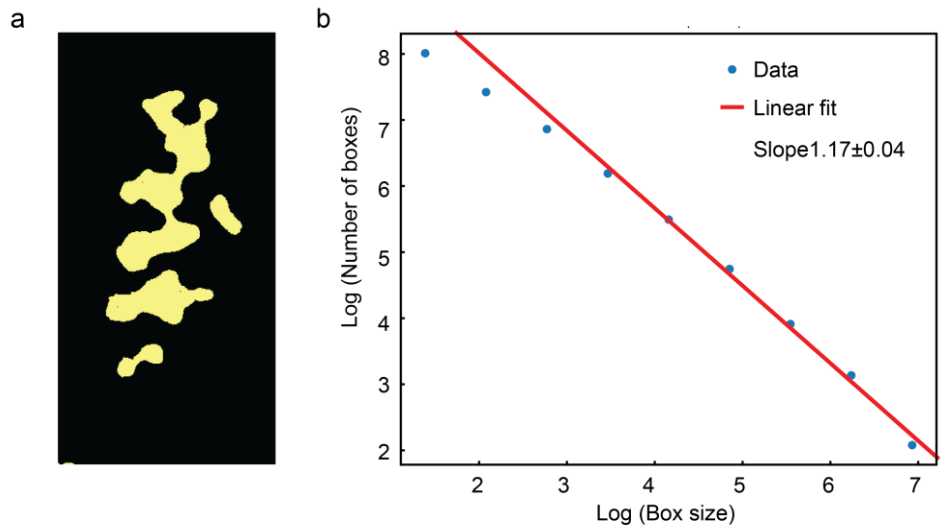

**Extended Data Fig. 2 An example of the fractal dimension calculation using the Minkovski-Bouligand method. (a)** Binary MFM image of the sample after illumination with 4 pulses. **(b)** Graph of the logarithm of box size plotted against the logarithm of box number with a linear fit showing a value for D = 1.17±0.04.

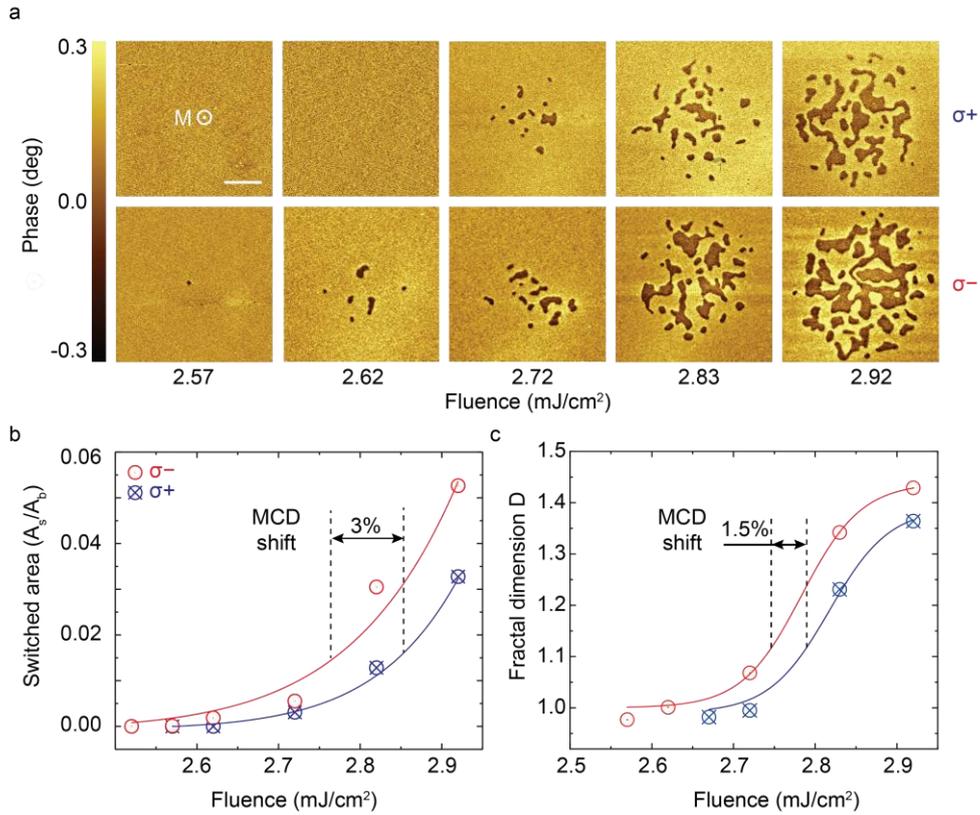

**Extended Data Fig. 3 Single-pulse nucleation of the SDN.** (a) MFM images of the single-pulse nucleated domains as a function of pulse fluence for left ($\sigma^-$) and right ($\sigma^+$) circularly polarized laser pulses. We observe that the threshold for small domains is $F_{\sigma^-}=2.57$ mJ/cm² for $\sigma^-$ pulses, while for $\sigma^+$ pulses, the critical threshold fluence is $F_{\sigma^+}=2.7$ mJ/cm². The difference between $F_{\sigma^+}$ and $F_{\sigma^-}$ confirms that the absorption of $\sigma^+$ and $\sigma^-$ pulses is inequivalent, which can be attributed to the MCD effect. The scale bar is 5 μm. (b-c) Switched area (b) and fractal dimension (c) are used as a function of fluence for the SDN, as shown in (a). Extracted an MCD-induced shift in $A_s/A_b$ as the function of fluence is approximately 3%. Fractal dimension $D$ also exhibits an MCD effect but with a slightly smaller induced shift of approximately 1.5%. Solid lines in (b,c) serve as a guide for the eye.

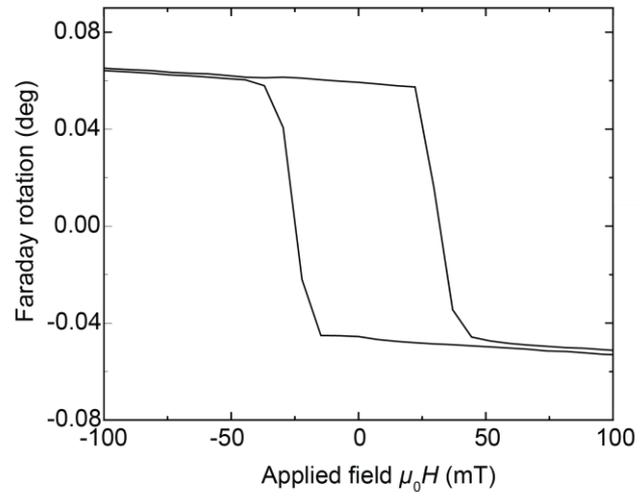

**Extended Data Fig. 4 Magnetic hysteresis loop for Pt(3 nm)/Co(0.6 nm)/Pt(3 nm).** The hysteresis loop is measured by observing the Faraday rotation in response to changes in an applied magnetic field.

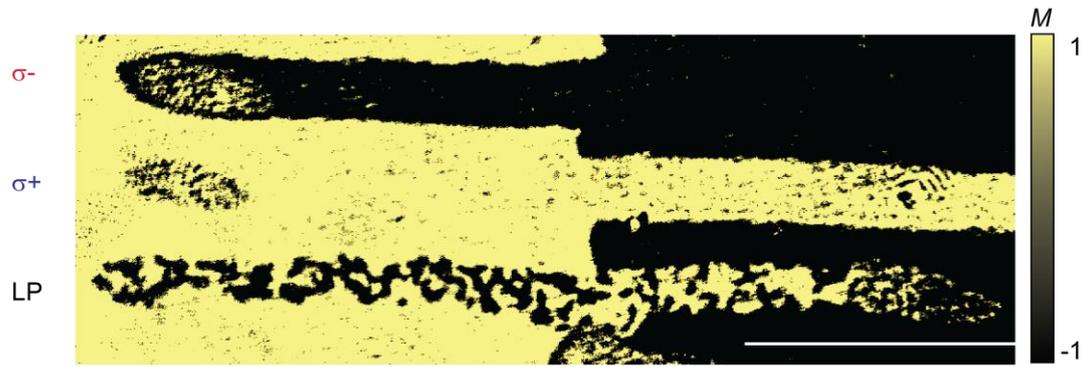

**Extended Data Fig. 5 Effect of the light helicity on the switching.** Magneto-optical image of two domains with $M=1$ and $M=-1$ after sweeping with a laser beam with $\sigma-$, $\sigma+$, and linearly polarized light (LP). Image shows that to switch magnetization from $M=1$ to $M=-1$ ($M=-1$ to $M=1$), $\sigma-$ ($\sigma+$) pulses are necessary. The scale bar is 100 $\mu$m

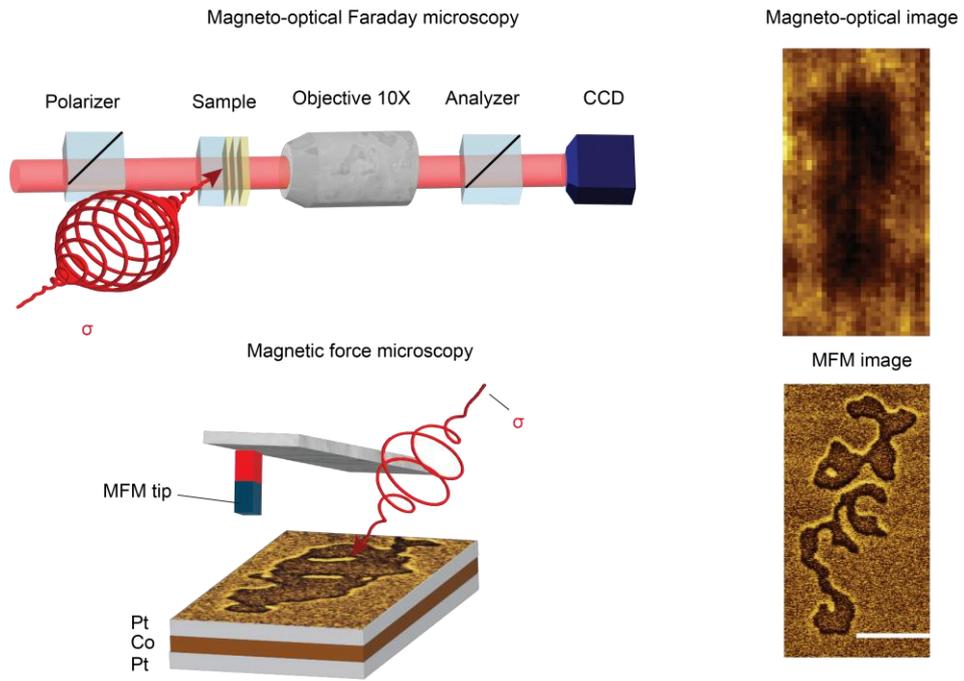

**Extended Data Fig. 6 Comparison of magneto-optical Faraday and magnetic force microscopy for studying laser-induced magnetic nanotextures.** The Scale bar is 5 $\mu$m

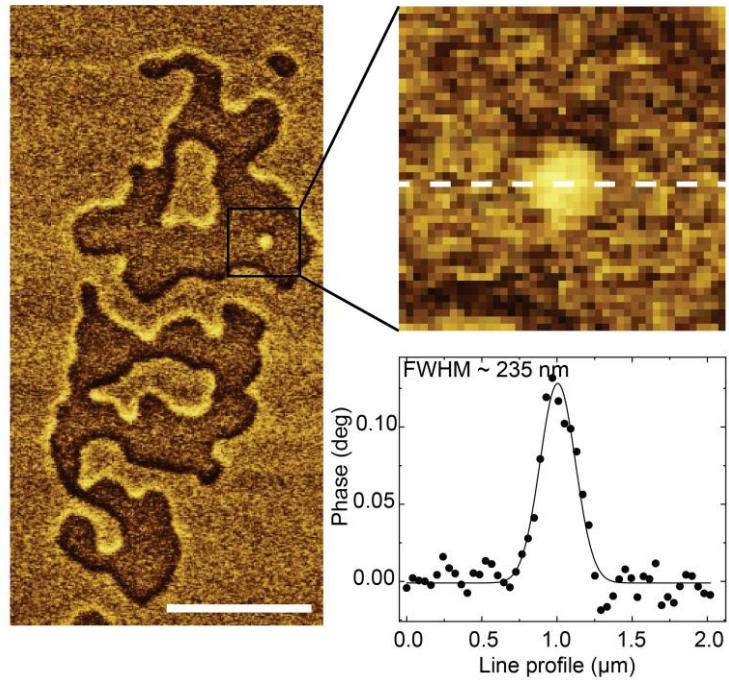

**Extended Data Fig. 7 Smallest laser-induced magnetic domain.** An MFM image of the laser-induced stochastic domain network was obtained after illumination with six pulses. The zoom-in panel shows the smallest magnetic domain. The line profile was fitted with a Gaussian distribution. The domain size is about 235 nm (Full width at half-maximum). The scale bar is 5 $\mu$m

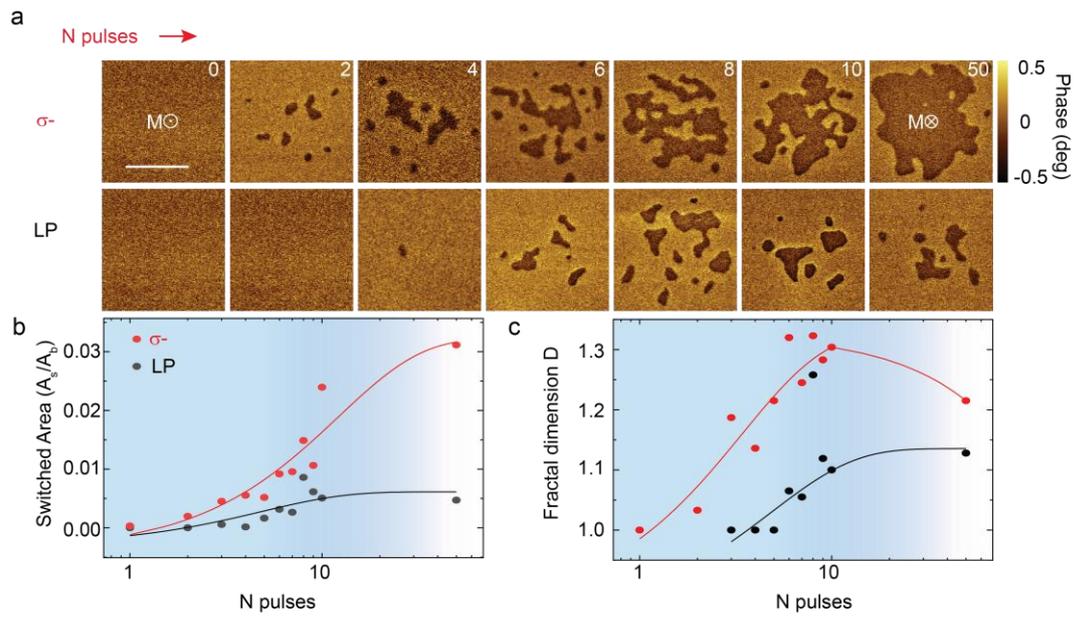

**Extended Data Fig. 8 Evolution of nanosized domains on the number of pulses for $\sigma^-$ and linearly polarized (LP) pulses**. **(a)** MFM images, **(b)** switched area, and **(c)** fractal dimension of the nucleated nanosized domains as a function of the number of $\sigma^-$ and LP pulses. Fluence was fixed at a value of about 2.5 mJ/cm$^2$. The light-blue region shows the SDN region. The scale bar is 5 $\mu$m.

# Supplementary Material: Laser-induced helicity and texture-dependent switching of nanoscale stochastic domains in a ferromagnetic film

Dinar Khusyainov[1*], Rein Liefferink[1], MengXing Na[1], Kammerbauer Fabian[2], Robert Frömter[2], Mathias Kläui[2], Dmitry Kozodaev[3], Nikolay Vovk[4], Rostislav Mikhaylovskiy[4], Dmytro Afanasiev[1], Alexey Kimel[1], Johan H. Mentink[1], Theo Rasing[1]

[1*]*Institute for Molecules and Materials, Radboud University, Nijmegen, 6525 AJ, Netherlands.*
[2]*Institute of Physics, Johannes Gutenberg University Mainz, Mainz, 55099, Germany.*
[3]*NT-MDT BV, Sutton 11A, Apeldoorn, 7327AB, Netherlands.*
[4]*Department of Physics, Lancaster University, Bailrigg, Lancaster, LA1 4YW, United Kingdom.*

*Corresponding author. E-mail: dinar.khusyainov@ru.nl;

## I. Supplementary Note

Using an analytical micromagnetic model, we estimate the switching energy barrier between $M = \pm 1$ of a cell in simulation for all possible configurations of neighbors. We consider a square domain of size $s = 200\ nm$ and its domain walls with the four fixed adjacent domains $M_{\text{neighb}} = \pm 1$ and calculate the micromagnetic energy barriers from the energy change under a 180-degree rotation. An example of a magnetization configuration during such rotation is given in Fig. S1. The micromagnetic energy is given by the out-of-plane uniaxial anisotropy and exchange energy

$$E = \iiint K_{\text{eff}}(\mathbf{M} \cdot \hat{\mathbf{z}})^2 \mathrm{d}x\mathrm{d}y\mathrm{d}z + \iiint A \left(\frac{\mathrm{d}\mathbf{M}}{\mathrm{d}x}\right)^2 \mathrm{d}x\mathrm{d}y\mathrm{d}z, \quad (1)$$

where $\mathbf{M}$ is the continuous magnetization field, $A$ is the exchange stiffness, and $K_{\text{eff}}$ is the effective out-of-plane anisotropy. Note that for simplicity, the dipole interaction is only taken into account as shape anisotropy to avoid long-range terms. It will be useful for the energy calculation to split the configuration into five regions, as indicated in Fig. S1.

The rotation is parameterized by the out-of-plane magnetization $-1 < m < 1$. For $m = -1$ to $m = 1$ domain walls (i.e., between antiparallel domains) in such a micromagnetic system, one finds an analytical solution for the domain wall profile [1] giving only non-zero values for the angle with the out-of-plane axis $\theta_{dw}(x) = 2\arctan\left(\exp(x/\Delta)\right)$, where $\Delta = \sqrt{A/K_{\text{eff}}}$ and $x$ is the distance perpendicular to the domain interface. To accommodate domain walls between $M_{\text{neighb}} = -1$ and arbitrary $-1 < m < 1$, we take an approximate profile

$$\theta_m(x) = 2\arctan\left(\exp\frac{x}{\Delta}\right)\frac{\arccos m}{\pi}. \quad (2)$$

For a wall from $M_{\text{neighb}} = 1$ to $m$, this becomes $-\theta_{-m}(x)$. For such profiles defined by only one angle $\theta(x)$, the micromagnetic energy density $\epsilon$ in the $yz$-plane can be written as

$$\epsilon = \int K_{\text{eff}}\sin^2\theta(x)\mathrm{d}x + \int A\left(\frac{\mathrm{d}\theta(x)}{\mathrm{d}x}\right)^2 \mathrm{d}x. \quad (3)$$

It is now easy to calculate the exchange energy density for the profile Eq. (2):

$$\epsilon_{\text{dw,exch}}(m) = \int_{-\infty}^{\infty} A\left(\frac{\mathrm{d}\theta(x)}{\mathrm{d}x}\right)^2 \mathrm{d}x = 2\sqrt{AK_{\text{eff}}} \cdot (\arccos(m)/\pi)^2. \quad (4)$$

Note that the profile Eq. (2), in general, never aligns with the anisotropy axis at positive x, so we must bind the extent of the wall to a finite value $x = t = 3\Delta < s/2$ (see Fig. S1). We obtain for the anisotropy energy density in the $yz$-plane

$$\epsilon_{\text{dw,ani}}(m) = \int_{-t}^{t} K_{\text{eff}}\sin^2(\theta_m(x))\mathrm{d}x \quad (5)$$

$$= K_{\text{eff}} \int_{-t}^{t} \sin^2\left(2\arctan\left(\exp(x/\Delta)\right) \cdot \frac{\arccos(m)}{\pi}\right) \mathrm{d}x, \quad (6)$$

for which no simple analytical solution can be found.

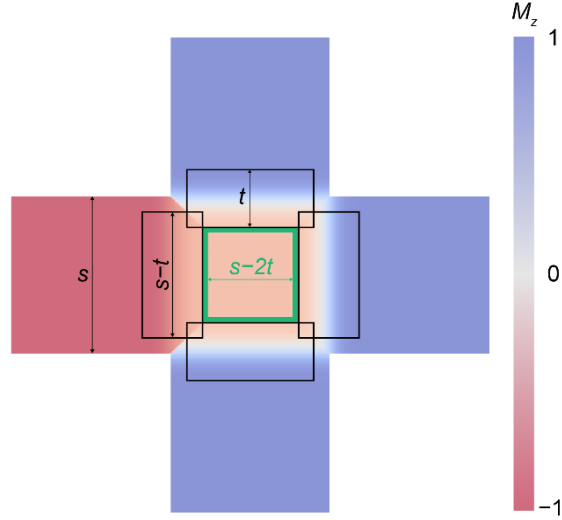

**1. Example of out-of-plane magnetization configuration under rotation of one model cell with a sketch of the subregions used for the energy calculation.** Note that the neighboring domains of the middle cell are fixed at $M_z = M_{neighb} = \pm 1$. Green and black frames represent the subregions used for the calculations. The subregions correspond to the homogeneous part of the cell (green) and the four domain walls (black).

These expressions hold for the profile from $M_{\text{neighb}} = -1$ to $m$, but we can write the energy of $M_{\text{neighb}} = 1$ to $m$ walls as $\epsilon_{dw,exch}(-m) + \epsilon_{dw,ani}(-m)$, since the $\theta$ only appears inside even functions. These walls stretch along the length $s$ of the domain and its walls, so we take their length as $s - t$. Then, the energy of the four walls is

$$E_{\text{dw}}(m) = d(s-t) \sum_{neighb\ i} \epsilon_{\text{dw,exch}}(\text{sgn}(-M_i)m) + \epsilon_{\text{dw,ani}}(\text{sgn}(-M_i)m), \tag{7}$$

where $d$ is the sample thickness.

The exchange energy of the homogeneous part of the domain between the domain walls clearly vanishes, but we still need to calculate its anisotropy energy. On either side, there is a half-domain wall of thickness $t$, so the remaining anisotropy energy is

$$E_{\text{domain,ani}}(m) = d(s-2t)^2 K_{\text{eff}}(1-m^2). \tag{8}$$

The switching energy barriers are now found from the difference between the peak of the energy during the rotation, i.e., the maximum of $E_{\text{dw}}(m) + E_{\text{domain,ani}}(m)$ for $-1 \leq m \leq 1$, and the initial energy $E_{\text{dw}}(M) + E_{\text{domain,ani}}(M)$. This gives an approximately linear dependence on the number of aligned (i.e., $MM_{\text{neighb}} = 1$) neighbors $n$

$$E_b(n) \approx E_0 + E_1 n. \tag{9}$$

We found the best correspondence of the domain shape as well as the fractal dimension for $A/K_{\text{eff}} = \Delta^2 = 285.7\,\text{nm}^2$, in which case the domain wall thickness is $d_{dw} = \pi\Delta = 53\,nm$. Fitting to equation Eq. (9) then leads to $E_0 = E_1$. Stochastic domain networks and cumulative switching could be achieved for other values $A/K_{\text{eff}}$ within an order magnitude, but the shapes and fractal dimensions deviated from the experimental results.